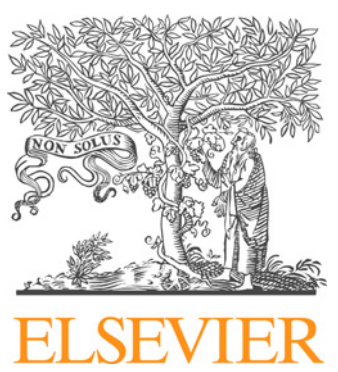

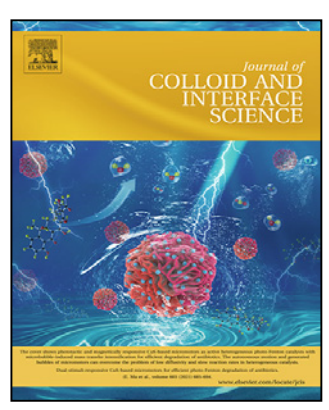

# Roughness induced rotational slowdown near the colloidal glass transition

Beybin Ilhan *, Frieder Mugele, Michael H.G. Duits *

*Physics of Complex Fluids, Faculty of Science and Technology, MESA+ Institute for Nanotechnology, University of Twente, Enschede 7500 AE, the Netherlands*

## GRAPHICAL ABSTRACT

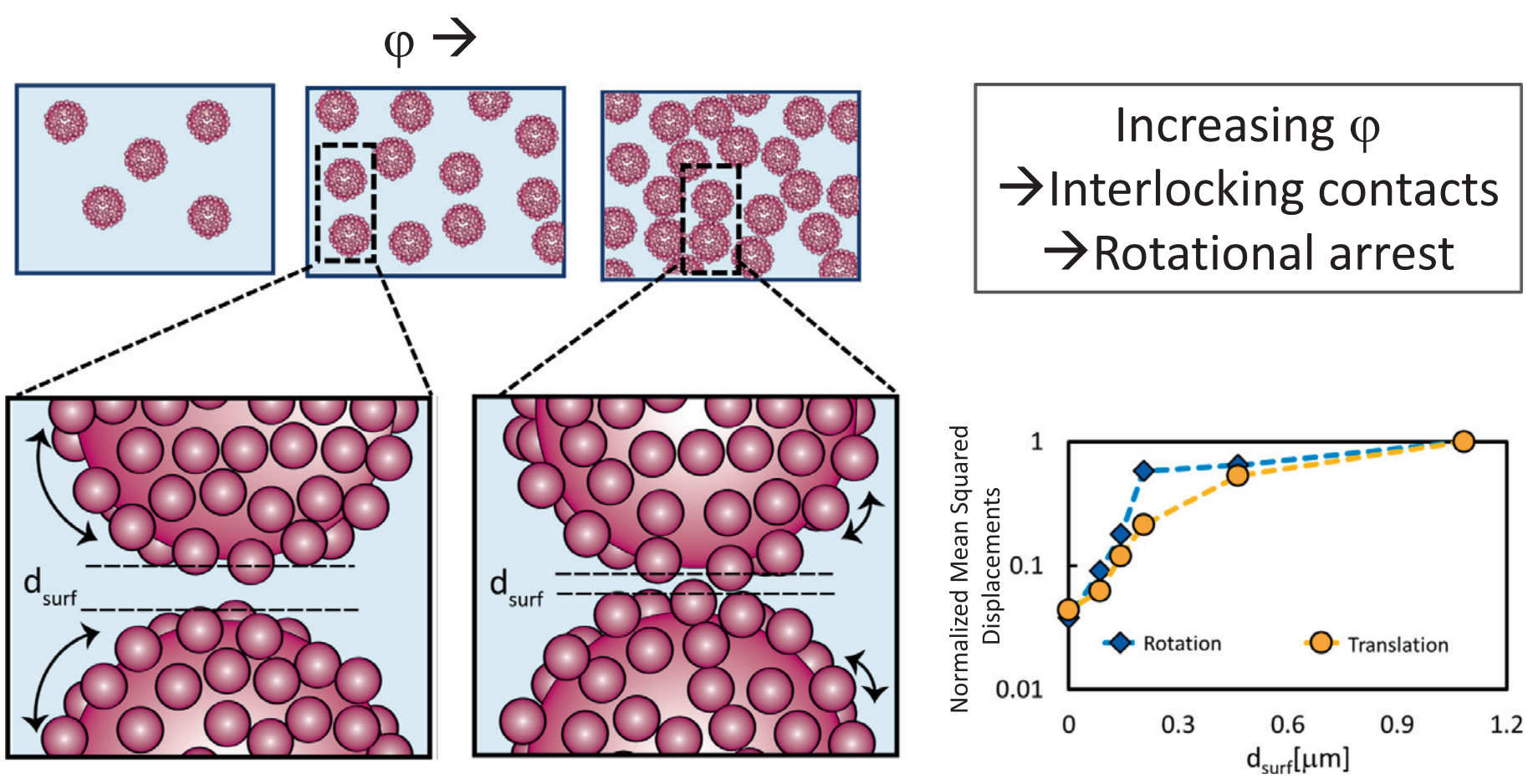

## ARTICLE INFO

## ABSTRACT

*Hypothesis:* In concentrated suspensions, the dynamics of colloids are strongly influenced by the shape and topographical surface characteristics of the particles. As the particles get into close proximity, surface roughness alters the translational and rotational Brownian motions in different ways. Eventually, the rotations will get frustrated due to geometric hindrance from interacting asperities.

*Experiments:* We use model raspberry-like colloids to study the effect of roughness on the translational and rotational dynamics. Using Confocal Scanning Laser Microscopy and particle tracking, we simultaneously resolve the two types of Brownian motion and obtain the corresponding Mean Squared Displacements for varying concentrations up to the maximum packing fraction.

*Findings:* Roughness not only lowers the concentration of the translational colloidal glass transition, but also generates a broad concentration range in which the rotational Brownian motion changes signature from high-amplitude diffusive to low-amplitude rattling. This hitherto not reported second glass transition for rough spherical colloids emerges when the particle intersurface distance becomes comparable to the roughness length scale. Our work provides a unifying understanding of the surface characteristics' effect on the rotational dynamics during glass formation and provides a microscopic foundation for many roughness-related macroscale phenomena in nature and technology.

* Corresponding authors.
*E-mail addresses:* b.ilhan@utwente.nl (B. Ilhan), m.h.g.duits@utwente.nl (M.H.G. Duits).

## 1. Introduction

Glass transitions in molecular systems remain a highly intriguing topic in materials science because of the spectacular changes in relaxation time and mechanical properties that occur precipitously upon reducing temperature [1–3]. Colloidal fluids are excellent model systems for studying glassy dynamics since not only they can reveal very detailed information about kinetic and equilibrium behaviors at experimentally accessible length and timescales on the single particle level [4], but also the volume fraction (rather than temperature) can be used as a control parameter [5,6].

Upon increasing the volume fraction, thus approaching the glass transition, colloidal dynamics slows down dramatically. This behavior is not only determined by the direction and the strength of the close range interactions [7], but also by the shape [8] and topographical surface characteristics of the particles. Although there have been substantial contributions to our understanding of the dynamics of glassy states using colloids with simple shapes (spheres, rods, ellipsoids) and smooth surfaces [9,10], arresting colloidal suspensions in nature and technology often consist of particles with highly irregular shape and roughness [11–14]. The current lack of studies into more complex particles also contrasts with the diversity and functionality of molecular systems and the recent advances in colloidal science. Many present-day colloids have a more complex architecture [15–17], yet understanding the dynamic interactions of these particles in dense suspensions and the extent of the coupling between the translational and rotational motions remains a standing issue.

Studying this uncharted territory requires experiments that resolve the thermal motions of individual particles. Remarkably, the few systems studied so far displayed opposing trends at the level of mean squared displacements (linear and angular): while colloidal tetramers suspended in a sea of spheres showed that rotations slow down more dramatically than translations [18], the opposite trend was reported for smooth Janus spheres [19]. Colloidal spheres with corrugated surfaces [20], showed that the rotations become slightly subdiffusive and the roughness was found to affect the rotations more strongly than the translations.

The current lack of an encompassing understanding is aggravated by the unclear role of surface inhomogeneities and topography as present on many practical colloids. Even minor deviations from smooth isotropic shapes and irregularities on the topography could alter the interplay between the rotational and translational dynamics if the particles get close enough [21]. Even though deviations from smooth surface topography have been shown to generate a broad diversity of macroscopic behaviors such as shear thickening and directed self-assembly [22,23], the interplay between microscopic dynamics and roughness has yet to be resolved.

In this work, we provide a unifying insight into colloidal suspensions' dynamics emerging due to a topography alteration to smooth spherical particles. By using our recently developed rotational probes with well-defined asperities [24] and time-resolved 3D confocal microscopy, we show that (unlike their smooth counterparts) rough particles undergo freezing in rotation as well as translation, resulting in two distinct glass transitions. Mapping rotational dynamics onto the number of particle–particle contacts supports the underlying physical picture that rotational slowdown can be driven by geometric frustration due to interlocking contacts among the particles.

## 2. Materials and methods

### 2.1. Materials

Deionized water (resistivity: 18 MΩ.cm) was obtained from a Millipore Synergy instrument. Core silica ($SiO_2$) particles (r = 1.05 μm) with amine ($NH_2$) surface modification were purchased from Microparticles GmbH. Two types of smaller $SiO_2$ particles (berries) were used: fluorescent spheres (sicastar-greenF, plain, r = 0.15 μm, labeled with Fluorescein-IsoThioCyanate (FITC)) were purchased from Micromod Partikeltechnologie GmbH, while plain $SiO_2$ particles were synthesized via the Stöber method [25]. TetraEthoxySilane, 25.0 wt% Ammonia, analytical grade Ethanol (99.9%, Emsol), 0.10 N Nitric Acid solution ($HNO_3$), glycerol (99.5%), fluorescein dye and LiCl were purchased from Sigma Aldrich. All purchased chemicals were used as received.

### 2.2. Synthesis and sample preparation

Our experimental system consists of all silica, raspberry-like colloidal particles made via the assembly of many small spheres onto a larger one. The synthesis of the particles is described in our previous work [24]. Briefly, raspberry particles were made by coating positively charged ($NH_2$-functionalized) $SiO_2$ spheres (*cores*, $r_c$ = 1.05 μm) with a dense layer of negatively charged small $SiO_2$ spheres (*berries*, $r_b$ = 0.150 μm) via electrostatic heteroaggregation in an aqueous medium. The berries cover the cores at high density yet still leave space for asperity contacts (Fig. 1.a). By supplementing the berries with 2–3% of equally sized fluorescent spheres prior to the reaction with the cores, raspberries with typically 4–6 fluorescent tracers were obtained, thereby creating the optical anisotropy required for measuring particle rotations. These fluorescent raspberries were mixed with non-fluorescent ones to facilitate the identification of individual raspberries (Fig. 1.b). All colloidal raspberries were coated with a 5–20 nm thin silica layer to ensure mechanical integrity. The surface roughness of the particles was quantified by analyzing Atomic Force Microscopy (AFM) height profiles for several raspberries. A spherical fit to the surface was subtracted from the height profile to isolate the roughness profile [24,26] (Fig. 1.c). After centrifugation the particles ($\rho_p$ = 1.68 g/mL) were dispersed in refractive index matching ($n_d$ = 1.44) mixture of water-glycerol (1:4 by wt., $\rho_s$ = 1.20 g/mL) to suppress Van der Waals interactions. 1 mM of fluorescein dye was added to the solvent to visualize and localize all raspberries regardless of fluorescent labels. Additionally, 1 mM LiCl was added to suppress the electrostatic interactions, giving a Debye length $\kappa^{-1}$ of ~ 8 nm. Ten target concentrations ranging from dilute to the glassy limit were prepared. More details of these procedures are given in Supplementary Material Section I.

### 2.3. Confocal Microscopy experiments

Samples were imaged using a confocal scanning laser microscope (CSLM) with a VisiTech 'VT-infinity3' scan unit, a 488 nm laser and 100x objective (1.49 N.A., oil immersion) connected to a Nikon Eclipse inverted microscope. 3D time series of ~ 61 × 60 × 8–12 $\mu m^3$ volumes were captured at fixed time intervals, chosen between 5.0 and 8.1 s. Since the solvent and the fluorescently labeled particles contain dyes with similar excitation wavelengths, we optimized the solvent dye concentration to achieve different intensity levels between the two. This allowed us to visualize all the particles regardless of the presence of fluorescent labels. Here the intensity inverted images were used to localize the centers of all raspberry particles, while the fluorescence of tracer berries was used for tracking the translations and rotations of labeled raspberries. The lowest focal plane was taken at least 10 μm above the glass bottom to avoid wall effects.

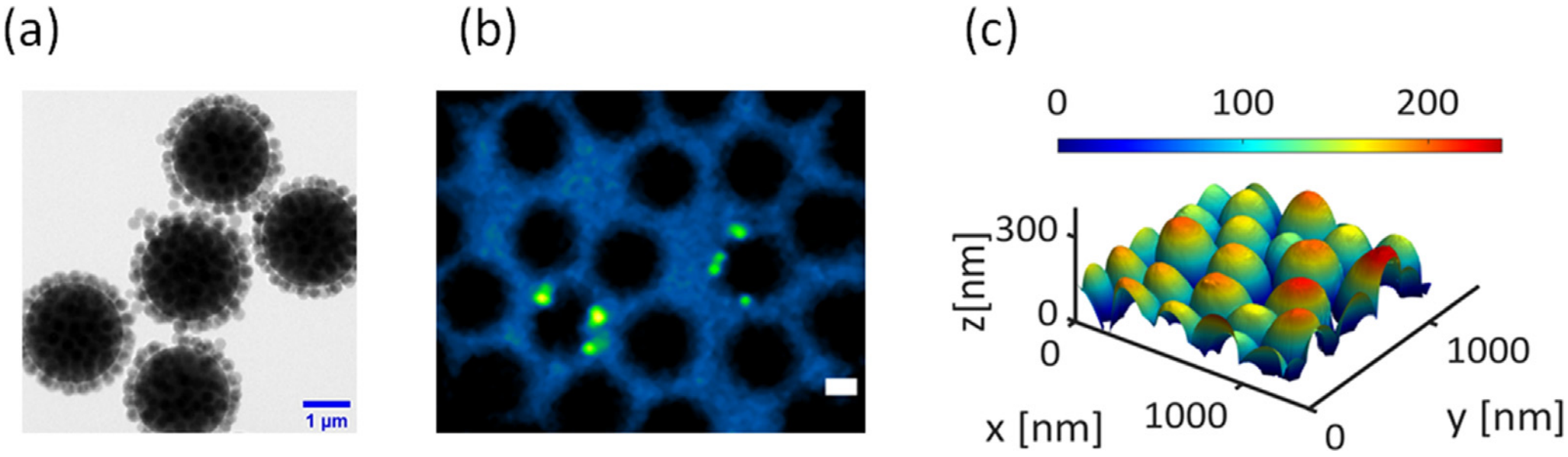

**Fig. 1.** (a) TEM image of raspberry particles, (b) 2D CSLM image showing the contrast between particles, tracer berries and solvent (false-colored, white scale bar 1 μm), c) Typical height profile of a raspberry (obtained from AFM measurements) after subtracting the curvature of the core particle.

### 2.4. Data analysis

3D localization of fluorescent berries and host particles was done using publicly available routines [27–29]. We used the method described in our previous work to extract and separate the tracer raspberries' rotations and translations [24]. By localizing at least 4 fluorescently labeled berries (per core), we determine a unique center-of-mass location by fitting a sphere encompassing the fluorescent berries' locations. Only the berries separated from each other by at least one berry size were analyzed to ensure correct assignment to a tracer raspberry in a crowded system. Lagtimes for which there were<50 mean squared (angular) displacement contributions were not taken into processing. After these filtering steps, we typically obtained ~ 20–30 (translational and rotational) trajectories lasting between 100 and 400 s. Translation of each tracer raspberry is then calculated from the time dependence of its center location. The latter location also provides an origin in a 3D Cartesian coordinate system that defines the tracer probe's orientation based on the angular positions of the fluorescent berries [30]. Rotational trajectories are then obtained by extracting the relative change in the angular positions in terms of rigid body transformations [24,30]. Extensive details on the data analysis are given in our previous work [24].

## 3. Results and discussion

### 3.1. Effect of roughness on translational and rotational dynamics

Centrifuging the suspensions (at a gravitational Peclet number of 1200) into CSLM cuvettes allowed us to directly assess the structure of the suspension, particle number density and translational Brownian motion at the highest achievable concentration in the bulk (i.e., more than10 μm from the bottom wall). This suspension displayed a lack of long-range order, while the translational Mean Squared Displacement showed no dependence on lagtime within the investigated time frame (see Fig. 2a), thus indicating glassy arrest. Multiplying the number density with the volume per raspberry, we find a volume fraction of 0.54 ± 0.02, i.e., significantly below the value of 0.62 ± 0.02, which we confirmed for a centrifuged suspension of monodisperse smooth spheres. This shift agrees with literature, where roughness [22,31] and friction [32,33] were found to lower the maximum packing fraction in the glassy state. Note that for raspberry particles, the definition of volume fraction is not entirely trivial. Therefore, when studying the glass transition, the volume excluded to other raspberries (rather than the volume excluded to solvent) should be used. Based on the dense assembly of small spheres in the raspberry shell (see inset Fig. 2), we estimate the effective raspberry volume to be 9.2 $\mu m^3$ (see also Supplementary Material). This corresponds to a smooth sphere with an effective radius $r_{eff}$ = 1.30 μm. The volume fraction dependent Mean Squared Displacements ($< \Delta r^2 >$) and Mean Squared Angular Displacements ($< \Delta\theta^2 >$) were measured at lagtimes that mostly ranged from 5 to 100 s, i.e., typically shorter than the characteristic time scales (obtained from Stokes-Einstein, Stokes-Einstein-Debye and < $\Delta r^2$ >, < $\Delta\theta^2$ > expressions for 1 $r_{eff}^2$ translation and 1 rad$^2$ rotation) for translational (88 s) and rotational (119 s) Brownian motion in the dilute limit.

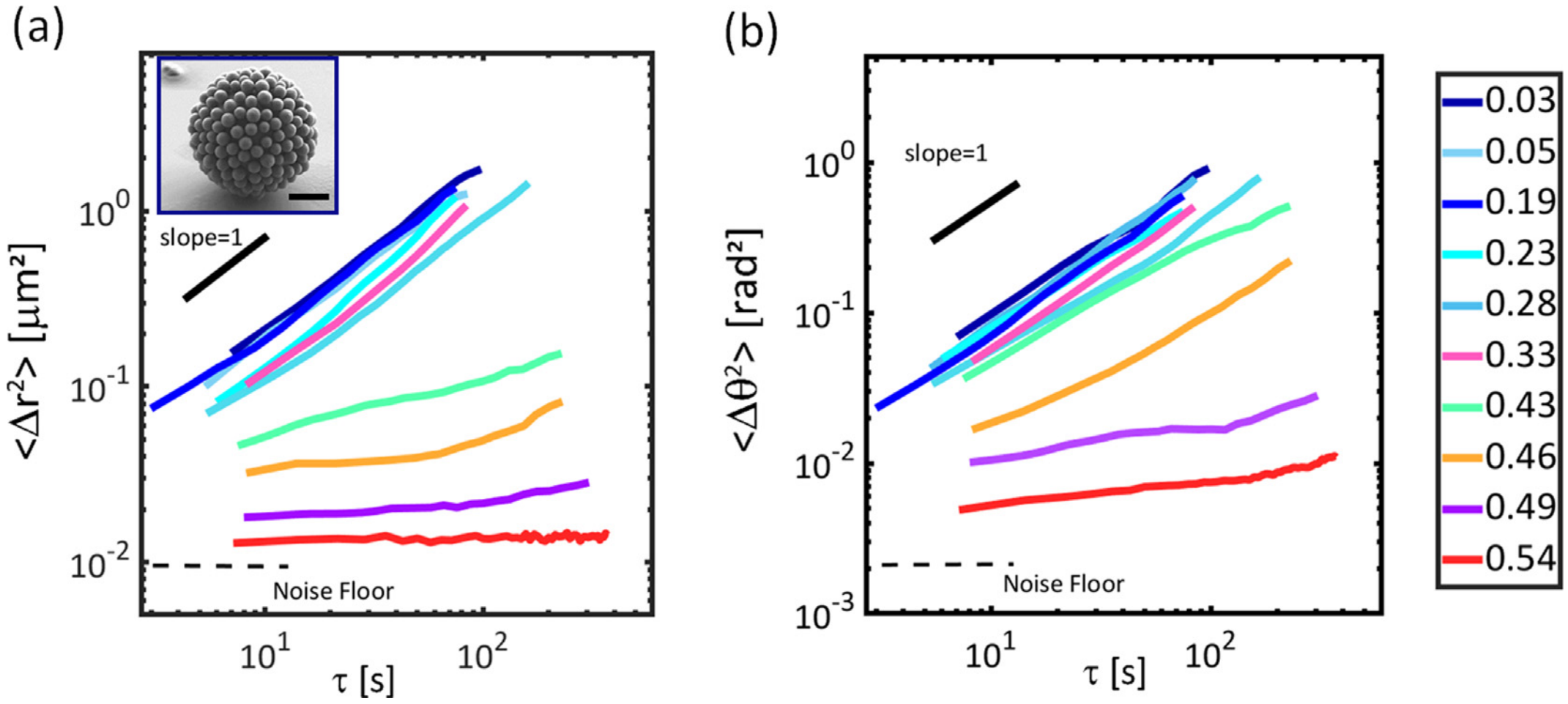

**Fig. 2.** (a,b) Volume fraction (φ) and lagtime (τ) dependent $< \Delta r^2 >$ and $< \Delta\theta^2 >$, based on the same image data per φ. Maximum packing is achieved at φ = 0.54. Whereas the translational diffusion plateaus for φ > 0.46, the rotational motion is sustained longer and becomes subdiffusive, ultimately plateauing at φ = 0.54. Noise floors are indicated with dashed lines. Inset: SEM image of a raspberry particle, scale bar = 1 μm.

For the translational dynamics (Fig. 2a), the typical diffusive signature of colloidal (smooth) hard spheres in the dilute limit is found to extend up to φ values as high as 0.33, while the amplitude of the < $\Delta r^2$ > undergoes a modest reduction. More drastic changes are found between φ = 0.33 and 0.43, where a subdiffusive regime emerges. On further increasing the concentration, < $\Delta r^2$ > shows plateaus for $\varphi \geq 0.46$, indicating that the particles become dynamically caged by their neighbors at the timescales of our measurement. The progressively lower < $\Delta r^2$ > plateaus at higher φ values reflect a further tightening of the cages, to the point that the motion amplitude approaches the noise floor. While this trend is qualitatively similar to that of smooth hard spheres [19], we recall that the maximum achievable volume fraction is now much lower at φ = 0.54.

The rotational mean square displacements (Fig. 2b) follow a similar trend with concentration, but the onset volume fractions for subdiffusive behavior and plateau formation are both significantly higher. Notably, at φ = 0.46, where the < $\Delta r^2$ > has already developed an initial plateau, < $\Delta\theta^2$ > still shows (almost) diffusive behavior. The transition to a < $\Delta\theta^2$ > plateau is observed at φ = 0.49 and continues up to φ = 0.54, where the < $\Delta\theta^2$ > levels are still significantly above the noise floor. This striking offset in the volume fractions, where the translation and the rotation get frozen, reveals -for the first time- two clearly separated glass transitions for rough spherical colloids. The distinct transition regimes in translational and rotational dynamics can further be seen from the volume fraction dependent powerlaw exponents, α, of the < $\Delta\theta^2$ > and < $\Delta r^2$ > (Fig. 3a) and from the ratio between the translational and rotational characteristic times ($\tau_R$, $\tau_T$) (Fig. 3b).

In Fig. 3a, the slopes are obtained from linear fits using lag times between 5 and 77 s, where the data are most accurate. Three distinct regimes are observed. In regime I, both translations and rotations are diffusive, suggesting that the particle concentration influences the Brownian motions via the macroscopic viscosity of the suspension. In regime II, the signature of < $\Delta r^2$ > steeply transitions into a subdiffusive one, while < $\Delta\theta^2$ > remains diffusive. This separation of dynamic signatures resembles that of smooth hard (Janus) spheres in the translationally glassy regime [19]. In regime III, both motions have become strongly subdiffusive. Ultimately both < $\Delta r^2$ > and < $\Delta\theta^2$ > reach (near) zero slopes, indicating that only 'rattling' motions are possible. Referring to Fig. 2b, we emphasize that the detected rotational motions at the highest concentration are well above the noise floor. We can thus conclude that the rotational rattling involves amplitudes up to ≈ 0.1 rad.

Fig. 3b illustrates another aspect of the dual glass transition with clear separation. Defining $\tau_R$ and $\tau_T$ as the typical times to rotate 1π rad$^2$, respectively translate $r_{eff}$, $\tau_R$ and $\tau_T$ were obtained from mean squared (angular) displacement functions after replacing them by the best fitting powerlaws and extrapolating where necessary. By plotting the ratio $\tau_T/\tau_R$ we can examine the relative slowdown of translation and rotation on increasing φ. As similar to the three distinct regimes observed in Fig. 3a, translations slow down drastically before the rotations do.

The question is, how to understand this influence of surface roughness on particle dynamics from a mechanistic point of view. Thanks to the refractive index matching solvent and added salt, the Van der Waals and electrostatic interactions are strongly suppressed. The dynamics of the colloidal raspberries should thus be dominated by excluded volume and hydrodynamic interactions. Similar to smooth hard spheres [34–36] at low to intermediate volume fractions ($\varphi \leq 0.33$), where the raspberry surfaces are separated by distances comparable to the particle radius, the hydrodynamic interactions between the overall particles should dominate. Bringing the particles closer together, the translational motions become constrained due to steric caging by surrounding particles. At still closer distances, contributions from the surface morphology start to play a role and eventually become dominant, causing the maximum achievable volume fraction to reduce to 0.54. Also, the differences in rotational dynamics between our raspberries and smooth hard spheres (SHS) could be explained using only hydrodynamic and steric interactions: at very high densities, the dynamics of SHS would be dominated by lubrication, while the rotational rattling behavior of the raspberries would also require steric interactions.

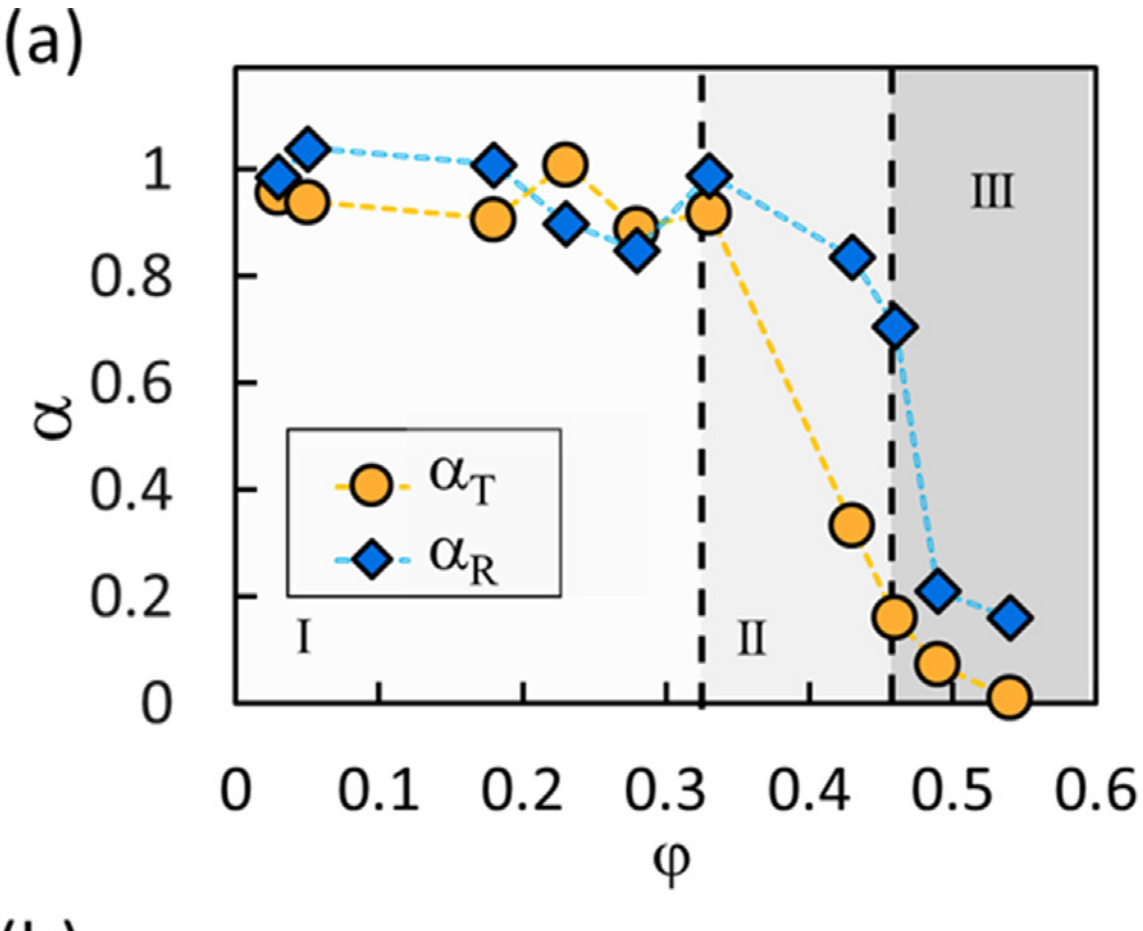

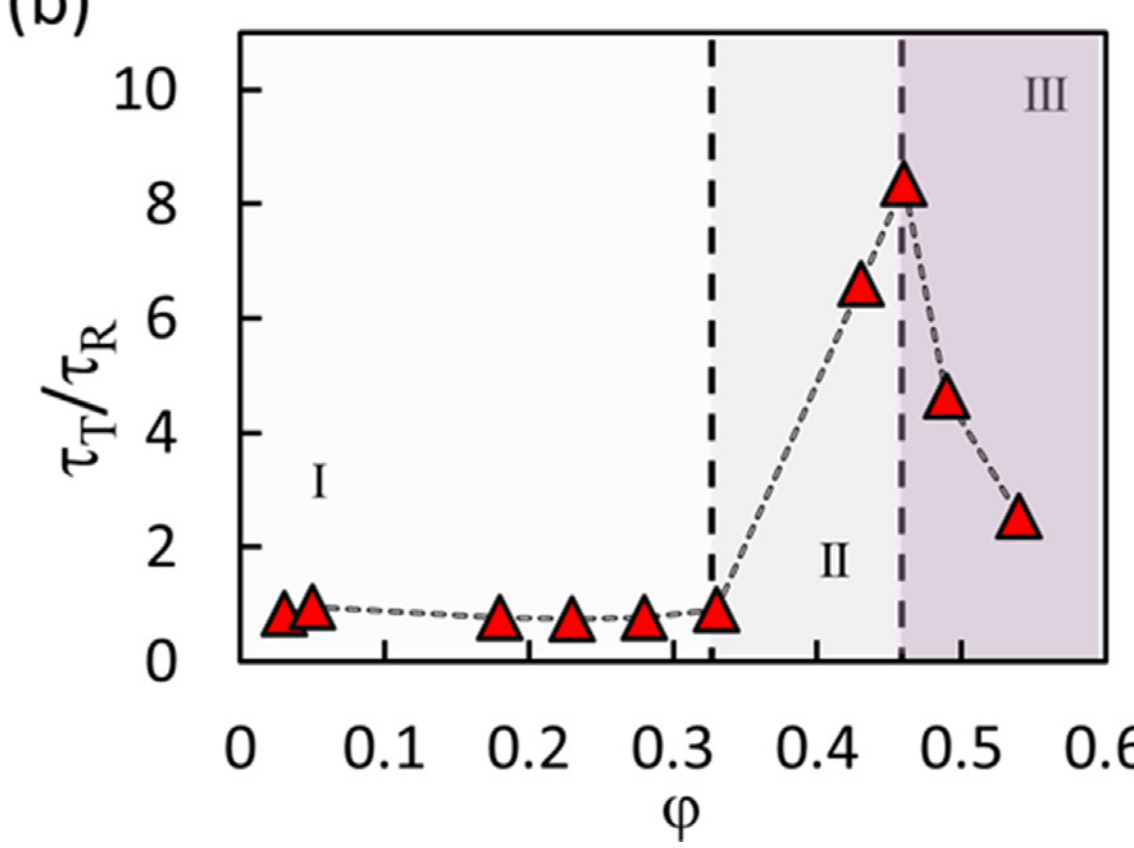

**Fig. 3.** Differences in the slowdown of the two types of Brownian motion as φ gets higher. (a) Log-log slope of the ensemble-averaged < $\Delta r^2$ > and < $\Delta\theta^2$ > versus lagtime. Slopes are obtained from linear fits using lag times between 5 and 77 s. (b) Ratio of translational to rotational characteristic time.

### *3.2. Physical picture of rotational slowdown*

Before this work, no rotational glass transition was reported for near-spherical colloids. Recently studied 'smooth' polymer latex (Janus) spheres showed a modest reduction of rotational motion while preserving the diffusive signature, even at very high volume fractions [19]. Polymer spheres with enhanced roughness due to surface cross-linking, displayed a transition into a subdiffusive rotational regime [20] without reaching the rotational rattling as in the present work. We, therefore, hypothesize that the distinct rotational glass transition can only be observed if the normal (amplitude) and lateral (spacing) roughness length scales are appropriately sized to allow interlocking, as sketched in Fig. 4a. When the raspberry particles are close enough, the asperities of one particle penetrating the void spaces of the other particle can generate interlocking configurations which restrict the angular degree of freedom to a narrow range.

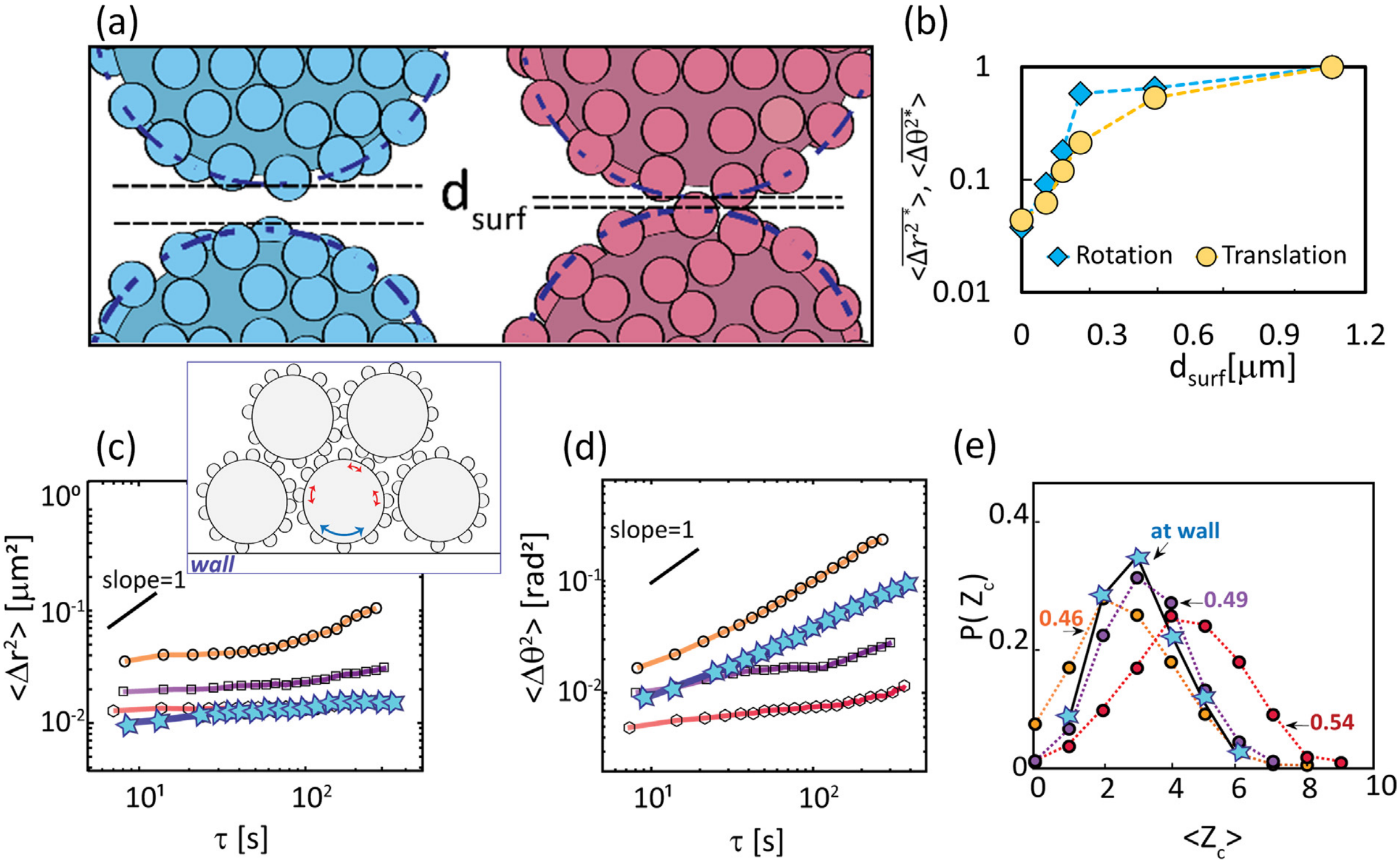

**Fig. 4.** (a) Artist impression of geometric interaction, with $d_{surf}$ the intersurface distance. Dashed lines around the raspberries represent the effective smooth radius $r_{eff}$ (b): Amplitudes of $<\Delta r^2>$ and $<\Delta\theta^2>$ at τ = 15 s, normalized by their values at φ = 0.19, plotted versus $d_{surf}$. Rotation decreases drastically for $d_{surf}$ < 0.2 μm. (c-d) $<\Delta r^2>$-and $<\Delta\theta^2>$ vs τ for φ = 0.46, 0.49, 0.54 and dense layer ($\varphi_{2D}$=0.59) of particles at the wall (respective colors and symbols: orange circle, purple square, red hexagon, blue star). Inset: Sketch of raspberries in contact with the wall. Red and blue arrows depict rotationally arrested and free contacts, respectively. (e) Contact number distribution for the same systems as in panel c-d.

To further reveal the link between the geometric frustration in rotational motion and particles' proximity, we plot the concentration-dependent $<\Delta\theta^2>$ and $<\Delta r^2>$ amplitudes against the average intersurface distance $d_{surf}$ in Fig. 4b. A simplistic estimate of $d_{surf}$ is obtained by assuming that all the particles are in direct contact at the maximum packing fraction (φ = 0.54), and lower volume fractions are reached via isotropic expansion on moving away from that point [37].

$$d_{surf} = 2r_{eff}[(\varphi_{max}/\varphi)^{1/3} - 1] \quad (1)$$

For comparison, we also calculate the average size of the cage $d_{cage}$ available to the raspberries for translational rattling. Considering that the apparent $<\Delta r^2>$ plateau is an indication of caging behavior for dense colloidal systems near the glass transition, we estimated $d_{cage}$ as $<\Delta r^2(\tau^*)>^{0.5}$ . Here, τ* serves as a representative lagtime and is chosen to be ~ 15 s (where the $<\Delta r^2>$ plateaus are significant for φ = 0.46–0.54).

Considering that neither distance measure is very precise, the correspondence between the calculated numbers in Table 1, where $d_{cage}$ should correspond to $2d_{surf}$, is satisfactory. Both Fig. 4b and Table 1 indicate that the rotational slowdown becomes prominent for distances $d_{surf}$ < 0.2 μm. This value is consistent with the amplitude of the roughness $d_{height}$, as extracted from AFM measurements showing a fairly dense and homogeneous distribution of berries with peak-to-valley height difference of 0.16–0.20 μm and a typical lateral length scale of ~ 330 nm. The agreement between these numbers underlines the proposed physical picture of interlocking

**Table 1**
Intersurface distance, cage size and coordination number of berries in the rotational glassy regime. $d_{surf}$ at $\varphi_{max}$ = 0.54 is taken as zero as all the particles are assumed to be in direct contact.

| φ | $d_{surf}$ [μm] | $d_{cage}$ [μm] | $<Z_c>$ |
|---|---|---|---|
| 0.43 | 0.21 | – | 2.1 |
| 0.46 | 0.14 | 0.19 | 2.7 |
| 0.49 | 0.09 | 0.14 | 3.3 |
| 0.54 | – | 0.12 | 4.3 |

as sketched in Fig. 4a and further corroborates our geometry based interpretation.

The emergence of (near) interlocking contacts for φ > 0.46 can be regarded as 'switching on' geometric frustration in the rotational degree of freedom. Interestingly, the same behavior can also be mapped by analyzing the number of contacts of each rough particle with other rough surfaces. The number of interlocking contacts, represented by the static coordination number, $Z_c$, is shown in Fig. 4e (as distributions) and included in Table 1 (as ensemble averages) for the volume fractions in the rotational glassy regime. Here $Z_c$ is calculated as the number of surrounding particles within a cutoff distance $d_{cut_off}$ (taken to be $2.2^*r_{eff}$ to account for polydispersity and localization errors). Increasing the concentration, $<Z_c>$-shifts to ≈4 at the maximum packing, where the system can be considered close to jamming. This value agrees well with the literature for athermal jammed packings, where $<Z_c>$ was found close to 4 ($Z_c$ = d + 1, with d as the dimensionality) for frictional spherical particles [33,38,39]. For comparison, we measured $<Z_c>$ for smooth hard spheres to be 5.9.

These outcomes raise the question, to what extent glass transitions could be controlled via the number of interlocking contacts per particle. Mixing rough spheres with smooth ones was already demonstrated as a possible means of controlling the mechanical properties of sheared glassy suspensions [22]. To examine this idea for quiescent fluids, we measured the dynamics of raspberries in contact with a smooth bottom wall (Fig. 4c, inset). This monolayer of particles forms a dense packing ($\varphi_{AF}$ = 0.59, 2D area fraction, $\varphi_{AF} > \varphi_{max}$ due to layering) while having a dense suspension of rough particles on top. Simplistically, the bottom wall could be regarded as a single smooth particle with a much larger radius. In this picture, the number of particle contacts gets reduced via i) the geometrical confinement (the wall allows only one contact while excluding other neighbors) and ii) the wall's smoothness. The $<\Delta r^2>$ and $<\Delta\theta^2>$ of this system are shown in Fig. 4c-d, along with the results for bulk φ = 0.46–0.54. It is evident that in this layer, the particle translations are strongly suppressed, with a $<\Delta r^2>$ similar to that of the bulk system at $\varphi_{max}$ = 0.54. No indications for a departure from plateau behavior (corresponding to translational cage-breaking) can be observed up to the longest lag time. In contrast, the $<\Delta\theta^2>$ for the dense layer of rough particles at the smooth wall still shows subdiffusive behavior, with significant amplitudes and a power law exponent $\alpha_r \approx 0.7$ that most closely resembles that of the bulk data at φ = 0.46. Also, a close resemblance in $<Z_c>$ for the 2 cases can be seen in Fig. 4.e.

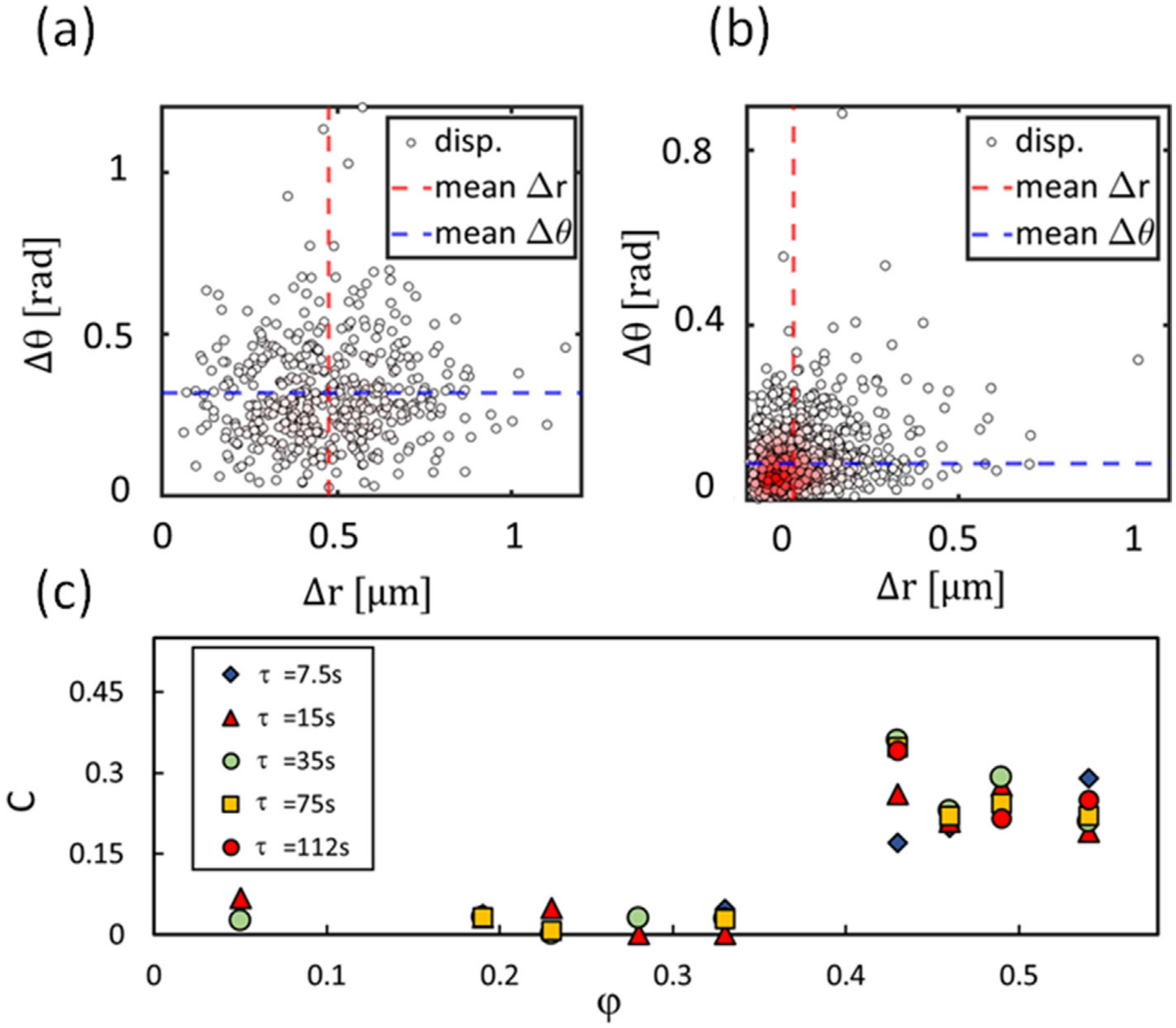

**Fig. 5.** Correlations between Δθ vs. Δr for τ = 15 s. Panels (a, b,) show scatter plots for volume fractions of 0.05, and 0.49 and dashed lines indicate the mean values. A color shade of red is given to guide the eye for density populations among the surrounding neighbors, i.e., light to dark: non-crowded to crowded. Panel (c) shows the correlation coefficient (see Eq (4)) obtained for all volume fractions and different lag times. While the motions are non-correlated at low φ, a significant coupling is observed for $0.43 \leq \varphi \leq 0.54$, where the rotational slowdown becomes prominent.

### 3.3. 3.3 correlations between translational and rotational displacements

The picture in which the translational Brownian motion influences the rotational diffusion by dynamically changing the contact number and distance between nearby raspberries also implies that the translational and rotational displacements of individual particles could be correlated. Here the reasoning is that a particle will have more vigorous Brownian motions when its instant location within the cage offers more space (and *vice versa*). We express the two types of mobilities via the scalar displacements: $\Delta\theta_i$ and $\Delta r_i$:

$$\Delta\theta_i(\tau) = |\boldsymbol{\theta_i}(t+\tau) - \boldsymbol{\theta_i}(t)| \quad (2)$$

$$\Delta r_i(\tau) = |\boldsymbol{r_i}(t+\tau) - \boldsymbol{r_i}(t)| \quad (3)$$

where bold symbols indicate (angular) location vectors, the subscript i labels trajectory segments with duration τ, and the vertical bars indicate absolute value notation. We use scalar quantities since only the amplitudes should be correlated with $d_{surf}$. The lagtime τ should ideally be taken as short as possible, since the coupling between $\Delta r_i$ and $\Delta\theta_i$ is governed by $d_{surf}$. Fig. 5a-b show typical scatter plots of $\Delta\theta$ vs. Δr for two representative volume fractions for a lag time of 15 s. Using the mean values $(\bar{\Delta r}, \bar{\Delta\theta})$ to define the origin, allows inspecting the plots for any asymmetry over the different quadrants. At φ = 0.05, where colloids are far apart, and both the $<\Delta r^2>$ and $<\Delta\theta^2>$ are diffusive, the point-cloud looks symmetric with respect to $(\bar{\Delta r}, \bar{\Delta\theta})$, indicating the absence of correlation between translational and rotational displacements. However, at φ = 0.49, where both translations and rotations are subdiffusive, more events where Δθ and Δr are simultaneously larger or smaller than their mean values indicate a positive correlation between 2 motions.

A more quantitative measure of the correlation is obtained from the lagtime-dependent correlation coefficient:

$$C(\tau) = \frac{\sum_{i=1}^{n}(\Delta r_i - \bar{\Delta r})(\Delta\theta_i - \bar{\Delta\theta})}{\sqrt{\sum_{i=1}^{n}\left(\Delta r_i - \bar{\Delta r}\right)^2 \sum_{i=1}^{n}\left(\Delta\theta_i - \bar{\Delta\theta}\right)^2}} \quad (4)$$

where all $\Delta\theta_i$ and $\Delta r_i$ depend on $\tau$ as before. Fig. 5c shows that regardless of the lagtime, C remains essentially zero up to $\varphi = 0.33$. For $\varphi = 0.43$ and higher, i.e., in the regime of strong rotational slowdown, significant correlations are found. These correlations' magnitudes are similar to those found for tetrahedral colloidal clusters in a supercooled colloidal liquid [18], where (somewhat similar to our case) geometric interactions are prominent. The absence of a clear dependence of C on lagtime in our data might be related to the finite number of observations (only the tracer berries contribute to the correlations). However, the different behavior of C in the two concentration regimes is explicit and underlines the picture that $\Delta\theta_i$and $\Delta r_i$ are correlated when the particles are in close proximity.

## 4. Conclusions and outlook

Using a novel type of rough colloidal probe spheres allowed us to simultaneously track translations and rotations with unprecedented precision. By studying the dynamics of our model rough colloidal system within a broad concentration range, we show that unlike their smooth counterparts [1], rough particles undergo glass transition in rotation as well as translation. On increasing the volume fraction, translational slowdown precedes the rotational one, indicating a distinct separation between two types of glass transition. This hitherto not reported second transition for isotropic-shaped particles reveals rotational rattling of the particles when the particle intersurface distance becomes comparable to the roughness amplitude. It is likely that the particles can then form interlocking contacts leading to geometric frustration. The obtained insights can pave the way for understanding the influence of other types of short-ranged colloidal interactions (electrostatic, Van der Waals, polymer-induced) on the rotational slowdown in glass formation [6,41]. This should lead to a better understanding of various roughness-related phenomena, including the role of geometric (rolling and sliding) friction in shear-thickening materials [13,39] and the processing of rough particulate systems in industrial applications. In addition, our results can provide a foundation for designing smart materials and superstructures with tunable directional mechanical properties [15,41] that can be obtained by controlling the asperity shape and number density of the surface protrusions.

## CRediT authorship contribution statement

**Beybin Ilhan:** Conceptualization, Software, Investigation, Visualization, Formal analysis, Data curation, Writing – original draft, Writing – review & editing. **Frieder Mugele:** Conceptualization, Supervision, Writing – review & editing. **Michael H.G. Duits:** Conceptualization, Data curation, Supervision, Writing – review & editing, Funding acquisition, Project administration.

## Declaration of Competing Interest

The authors declare that they have no known competing financial interests or personal relationships that could have appeared to influence the work reported in this paper.

## Acknowledgements

We thank Dirk van den Ende and Stefan Luding for helpful discussions. This work was financially supported by NWO-CW (ECHO grant 712.016.004).

## Appendix A. Supplementary material

Supplementary data to this article can be found online at https://doi.org/10.1016/j.jcis.2021.08.212.